# *Viable route towards large-area two dimensional MoS$_2$ using magnetron sputtering*

Hassana Samassekou[1], Asma Alkabsh[1], Milinda Wasala[1], Miller Eaton[1], Aaron Walber[1], Andrew Walker[1], Olli Pitkänen[2], Krisztian Kordas[2], Saikat Talapatra[1], Thushari Jayasekera[1] and Dipanjan Mazumdar[1,†]

1. *Department of Physics, Southern Illinois University, Carbondale IL 62901*
2. *Microelectronics Research Unit, Faculty of Information Technology and Electrical Engineering, University of Oulu, Finland*

## ABSTRACT

Structural, interfacial, optical, and transport properties of large-area MoS$_2$ ultra-thin films on BN-buffered silicon substrates fabricated using magnetron sputtering are investigated. A relatively simple growth strategy is demonstrated here that simultaneously promotes superior interfacial and bulk MoS$_2$ properties. Few layers of MoS$_2$ are established using X-ray reflectivity, diffraction, ellipsometry, and Raman spectroscopy measurements. Layer-specific modeling of optical constants shows very good agreement with first-principles calculations. Conductivity measurements reveal that few-layer MoS$_2$ films are more conducting than many-layer films. Photo-conductivity measurements reveal that the sputter deposited MoS$_2$ films compare favorably with other large-area methods. Our work illustrates that sputtering is a viable route for large-area device applications using transition metal dichalcogenides.

---

† Corresponding author: dmazumdar@siu.edu

# INTRODUCTION

During the last decade, layered materials have caused a paradigm-shift in our understanding of the fundamental properties of nanomaterials and opened up new technological possibilities. The discovery of single-layer graphene [1, 2] and transition metal dichalcogenides (TMDs) [3, 4], initially through mechanical exfoliation, have sparked a series of high profile discoveries that impact numerous electronic and optoelectronic areas [5-7], and opened up potentially new areas [8, 9].

TMDs offer many advantages because of their unique and tunable electronic properties. $MoS_2$, one of the most investigated TMD in recent years, changes from an indirect (1.3 eV) to a direct (1.8 eV) gap system as it is shrunk to a monolayer [10]. Other TMDs such as $MoSe_2$ [11], $WS_2$, $WSe_2$ [12] show similar properties. It is not a surprise that such materials have demonstrated many high performance devices using single or bilayer $MoS_2$ such as Field effect transistor [13], Photodetector [14, 15], Memory [16], Integrated circuit [17] to name a few.

Growth of large-area $MoS_2$ has been demonstrated primarily using chemical vapor deposition (CVD) technique [18-21]. Vacuum-based methods such as pulsed laser deposition (PLD) [22-24], and magnetron sputtering [25] have also displayed promise. Also many applications necessitates that the TMDs be compatible with a wide variety of substrates, apart from large-area growth. Previous studies have shown that optical properties of $MoS_2$ and other TMDs are significantly influenced by the underlying substrate [26-28]. In particular, in a very recent study [28], using mechanically exfoliated $MoS_2$ and h-BN flakes, it was shown that an atomically thin buffer layer of h-BN protects a range of key properties of monolayer $MoS_2$.

Motivated by such developments, we examine the properties of large-area $MoS_2$ thin films down to approximately four layers deposited on thin amorphous Boron nitride buffered silicon substrates grown using r.f magnetron sputtering. The focus of this work is to investigate simultaneously interface and bulk structural properties leveraging primarily upon established large-area thin-film characterization techniques such as X-ray reflectivity and diffraction. Spectroscopic ellipsometry, transport, and first-principles calculations are employed to investigate the electronic properties of the large-area films. Previously we have demonstrated growth of many-layer $MoS_2$ films with well-defined Raman peaks and photo-responsivity [29]. Here we extend and deepen the scope of the previous work. What distinguishes our work is that we have developed a relatively simple strategy to grow high quality few-layer $MoS_2$ that mainly comprises of post-deposition annealing after a low temperature growth. Combined with ellipsometry and Raman characterization we confirm that it is possible to deposit highly uniform, large area few layers (~3-4 layers) of $MoS_2$. Transport measurements reveal that the conductivity of few-layer films is higher that the many-layer films deposited with our method. We also show that electrical conductivity of devices fabricated using these deposited films show very consistent results, indicating that the process presented here can lead to large area, few layer TMD layers with reliable physical properties.

## METHODS

Boron nitride (BN) and $MoS_2$ layers were grown using commercially available stoichiometric targets and r.f sputtered in a high vacuum magnetron sputter system (base pressure $4x10^{-9}$ Torr) under different deposition conditions. Particularly, growth temperature (room temperature to 400C), and post-deposition annealing temperature (300-750C) were varied. Deposition of amorphous BN was carried out in a 1:10 $N_2$/Ar mixture to produce stoichiometric films. Based on our prior result we concluded that high temperature growth (300C or over) could lead to unwanted interfacial layers and disordered $MoS_2$ [29]. As we show here, an appropriate post-deposition annealing treatment following a low temperature growth (best at room temperature) leads to sharp interfaces and crystalline $MoS_2$

X-ray reflectivity (XRR) and X-ray diffraction (XRD) patterns were evaluated using a high resolution Rigaku Smart Lab X-Ray diffractometer equipped with a channel-cut Ge (220) crystal to produce a highly monochromatic K-alpha radiation. XRR data was analyzed using GenX software [30] that implements the Parratt formalism of multi-layers [31]. Optical constants such as the refractive index (n, k) were measured in the 0.8-4.2 eV range using variable angle spectroscopic ellipsometry (J.A. Wollam WVASE and M2000V) and modeled using the VWASE32 software to extract layer-specific properties. Raman measurements were performed with Nanophoton Raman-11 spectrometer using a 532 nm laser. Both large-area and single-spot measurements were performed to verify the properties of $MoS_2$ films.

## RESULTS AND DISCUSSION

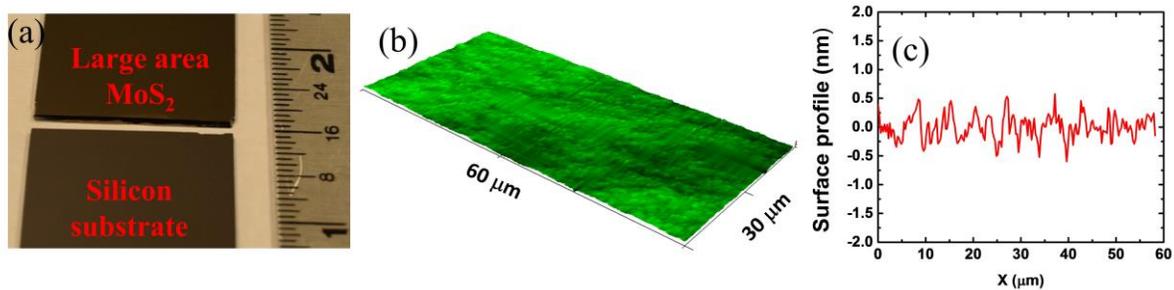

*Figure 1 (a) Optical image of an optimized large-area 20nm $MoS_2$ sample grown on BN-buffered Silicon substrate using sputter deposition method. A bare silicon substrate is also shown in comparison to highlight the color contrast. (b) A 60×30 μm AFM scan of a 20nm $MoS_2$ film showing smooth surface morphology. (c) Typical surface profile (line scan) of the AFM data shown in (b). The average peak-to valley variation is less than 1nm.*

In Figure 1a we show the optical image of an optimal 20 nm $MoS_2$ film deposited on BN-buffered Silicon substrate. In comparison to a bare silicon substrate the color contrast is distinct. The thickness homogeneity of the sample is primarily limited by the size of sputtering target (2 inch diameter in our sputtering system), and we expect our films to be very uniform over 1×1 inch area. In fig 1b, we show a large atomic force microscopy (AFM) scan of a 10-20nm (many-layer) $MoS_2$ film. As evident from the nearly uniform color, and also from the surface profile data (fig 1c), our samples are extremely smooth with a root-mean-square roughness of less than 0.25 nm. Thinner, few-layer samples are smoother. This is typically of a three-dimensional growth mode that is expected for the low temperature synthesis employed here (see Fig S3 in supplementary).

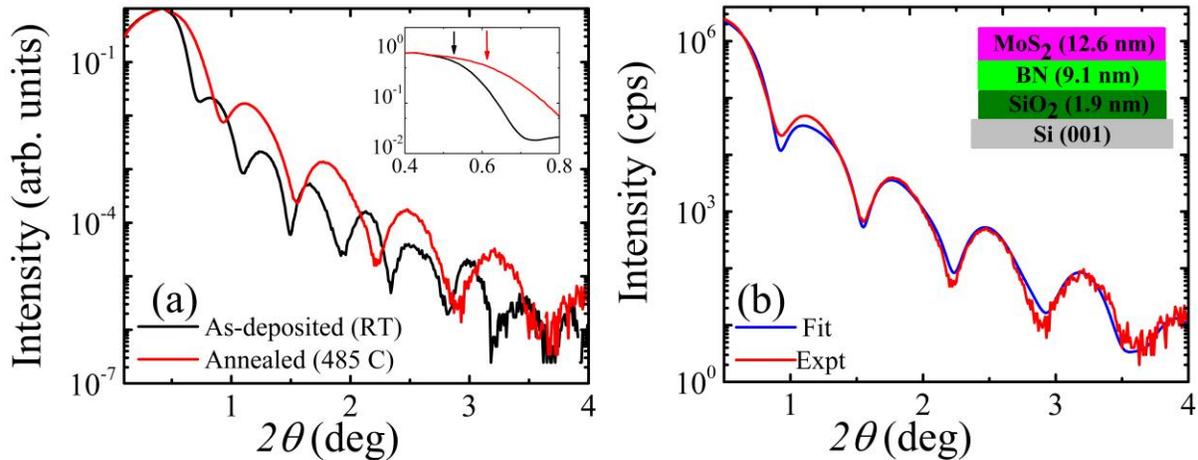

Figure 2. (a) X-ray reflectivity curve of the BN-$MoS_2$ heterostructure grown on Si (001) substrate before (black), and after (red) annealing. Inset of 1(a) shows the change in critical density in the two cases. (b) The optimally annealed sample (485 C) is fitted to obtain the layer-specific thickness, roughness, and density. The thickness values of the layers are listed in the inset schematic diagram.

In Figure 2 (a), we show the X-ray reflectivity (XRR) data for a many-layer $MoS_2$ film grown on sputtered a-BN at room temperature (black line, labeled "RT") and after a 2hr-485C annealing treatment (red line, labeled "485C"). Clear fringes are observed in both cases up to $2\theta = 4°$. This is a qualitative feature of sharp interfaces over large area. However to proceed further, a quantitative analysis is needed [30, 31]. Various relevant information can be extracted from XRR data such as thickness, large-area roughness, and density values of each individual layer; and, the values for both conditions are outlined in Table S1. The fit data for the annealed sample is shown in Fig. 2 (b) along with the extracted thickness values for the different layers in the inset. Let us discuss the XRR features in some detail. The critical angle, $\theta_c$, the angle at which X-rays start to penetrate the sample is related to the density of the underlying layer(s). Higher $\theta_c$ indicates higher density which is realized for the annealed (485C) sample as shown in the inset of Fig. 2. The extracted $MoS_2$ density (0.018 g/cm$^3$) after annealing is very close to the bulk value (0.019 g/cm$^3$). Our analysis also finds a large 7 nm (35%) change in $MoS_2$ thickness upon annealing that can be qualitatively inferred from the changing periodicity of the thickness oscillations. This is suggestive of major structural changes and consistent with crystallization of the $MoS_2$ layer, as also evidenced in our XRD, Raman and optics data (discussed later). No additional interfacial layer was required to obtain a good fit apart from a native $SiO_2$ layer. This is in contrast to a high temperature growth method that we have reported earlier [29]. Therefore, we can conclude that a low temperature growth with appropriate post-deposition annealing promotes desirable features for $MoS_2$.

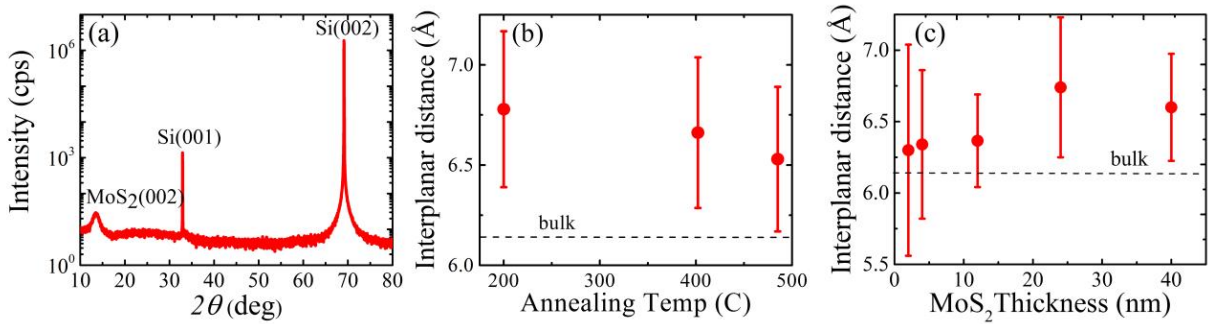

*Figure 3. (a) X-ray diffraction pattern of many-layer MoS$_2$ on BN-buffered Si substrate. (b) Variation of distance between MoS$_2$ layer as a function of annealing temperature. (c) The measured out-of-plane interplanar distance vs film thickness for optimally annealed samples.*

In Figure 3 (a), we show the $\theta$–$2\theta$ scan for a thick, 40 nm $MoS_2$ film. Clear diffraction peaks are obtained for Silicon and the (002) $MoS_2$ layer as indicated. We have verified that the low angle peak is from $MoS_2$ and not BN by measuring separate BN thin film samples. The effect of annealing temperature on the inter-planar distance of $MoS_2$ is shown in Fig. 3 (b). The lattice parameter gradually decreases at higher temperatures which is consistent with the expected strain-relaxation effect, but still somewhat larger than the bulk value (dashed green line). The peak intensity also increases substantially with increase in temperature (Fig. S1), implying better crystal quality with annealing. However, beyond 485C, multiple phases appear in the XRD dada (Fig. S1) implying that $MoS_2$ is chemically unstable beyond a certain high temperature.

Based on our XRR thickness calibration values (and ellipsometry), we fabricated a series of $MoS_2$ samples down to approximately 2.5 nm. The evolution of the interplanar distance as a function of thickness is plotted in Fig. 3 (c) for the samples shown in Fig. S2 and Table S2. Our result indicates that the $MoS_2$ layer sharply transitions to bulk-like lattice parameter values for less than 10nm thickness. This could imply that thicker (>10nm) films have a high degree of disorder in the form of defects and vacancies, a finding that is also supported by Raman measurements, but contrary to our initial expectation. From now on, we shall primarily discuss the properties of the few layer $MoS_2$ sample (labeled FL-$MoS_2$) of thickness ~2.5 nm that corresponds to 4 layers.

We have performed additional characterization using Raman spectroscopy in order to verify thickness and properties of our $MoS_2$ films (Figure 4).

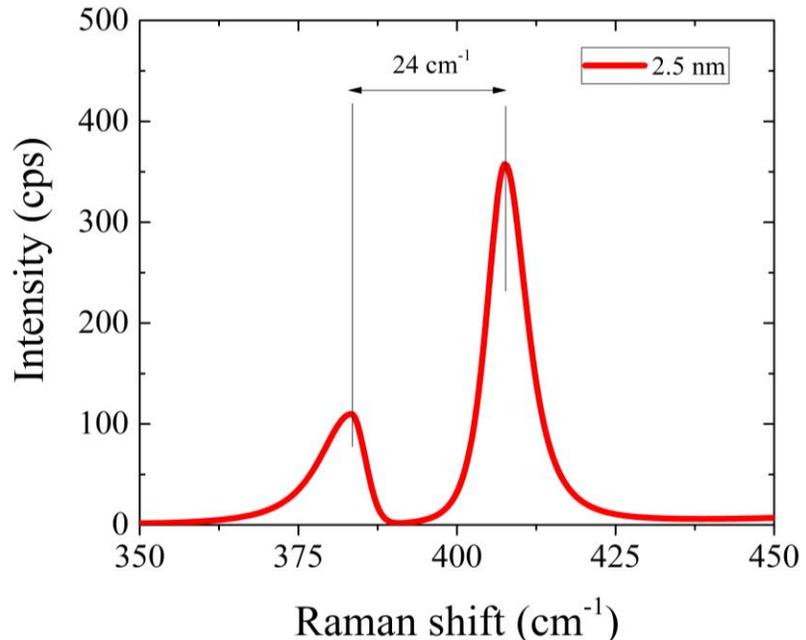

*Figure 4. Raman spectra of the FL-$MoS_2$ sample (~2.5 nm) showing the characteristic $E^1_{2g}$ and $A_{1g}$ vibrational modes of $MoS_2$*

Raman spectroscopy is a strong tool, which is widely used in order to determine the vibrational modes of layered chalchogenides. The values obtained for various vibrational modes are akin to the layer numbers and hence can be used to determine the number of layers. For example, in case of MoS$_2$ two predominant vibrational modes, one at ~407 cm$^{-1}$ another at ~383 cm$^{-1}$ belong to the out of plane A$_{1g}$ mode and the in plane E$^1_{2g}$ mode respectively. The frequency difference ($\Delta\omega$) between the A$_{1g}$ and the E$^1_{2g}$ modes is a good indicator of number of layers present in MoS$_2$. Typically $\Delta\omega$ ~26 cm$^{-1}$ corresponds to bulk MoS$_2$, whereas $\Delta\omega$ ~18–19 cm$^{-1}$ corresponds to monolayer of MoS$_2$. We have consistently obtained a value of $\Delta\omega$ ~ 24cm$^{-1}$ for FL-MoS$_2$ samples as shown in fig. 4 indicating that they are indeed closer to 4 layers. We have also seen that $\Delta\omega$ values slowly approaches to ~26cm$^{-1}$ for thick samples consistent with our past results [29].

Another indicator of film quality is the full width at half maxima (FWHM) of the observed vibration modes. FMHM values for the sputtered FL-MoS$_2$ film is compared with literature values in Table S3. Typical values for exfoliated MoS$_2$ and CVD grown films is ~2-6 cm$^{-1}$ for the E$^1_{2g}$ mode, and 4-6 cm$^{-1}$ for the A$_{1g}$ mode. In comparison, our 4-layer films have slightly higher values of 8.5 and 7.4 cm$^{-1}$ respectively. Apart from vacancy induced disorder, roughness from the BN underlayer could also contribute to the observed FWHM. This will be tested in future studies. Another observation is that the many-layer films show much higher FWHM (10-15 cm$^{-1}$) that is consistent with our earlier inference that thicker films (>10nm) are more disordered compared to few-layer films. This also has important implications on the transport properties of few-layer films as discussed later.

To check the robustness of the Raman modes with respect to device processing, we have characterized the deposited films before and after a photo-lithography process. The consistency of the Raman data obtained from multiple areas of many-layer sample (see figure S4) on the films before and after photo lithography process indicates the robustness of these films. From the measurements it is clear that the photolithography process does not destroy the integrity of the deposited films. Similar Raman results were also obtained from the few-layer MoS$_2$ samples (data not shown).

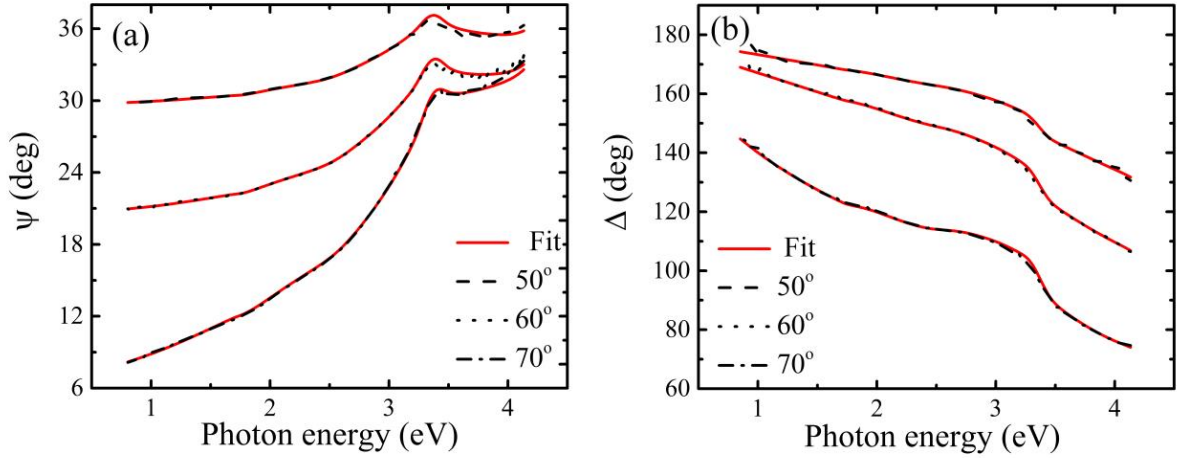

*Figure 5. (a) Experimental and fitted Psi ($\psi$) and (b) Delta ($\Delta$) for the FL-MoS$_2$ sample for incidenct angle of 50,60,70 degrees. Fit of these data provided a thickness of 2.6 nm.*

The optical properties of sputter-grown MoS$_2$ samples of various thickness were investigated using spectroscopic elipsometry (SE) in the 0.8-4.2 eV range. Ellipsometry is a popular, non-destructive technique that can provide thickness and optical constant values of thin films by measuring the ratio of p- and s- reflectance. This ratio is depicted by a complex number ($\tilde{\rho}$) given by

$$\tilde{\rho} = \frac{\widetilde{R_p}}{\widetilde{R_s}} = \tan\Psi \; e^{i\Delta} = \frac{\left(2\pi\tilde{n}\frac{d}{\lambda} \cos\phi_I\right)_p}{\left(2\pi\tilde{n}\frac{d}{\lambda} \cos\phi_I\right)_s}$$

where $\widetilde{R_p}$ and $\widetilde{R_s}$ are Fresnel p-and s-polarized reflection coeffficients, $\psi$ is amplitude ratio, $\Delta$ is the phase shift, d is the film thickness, $\tilde{n}$ is the complex refractive index, $\lambda$ is the wave length, and $\phi_I$ is the angle of incidence. The experimental $\psi$ and $\Delta$ data obtained at 50, 60, and 70 degrees for the FL-MoS2 sample is shown in Figure 5 (dashed black). To extract the optical constants, the data was fitted (red) to an emphirical four-layer heterostructure model containing semi-infinite Si, a native $SiO_2$ (~2nm) layer, BN, and $MoS_2$. The $MoS_2$ layer was modeled using a series of Kramer-

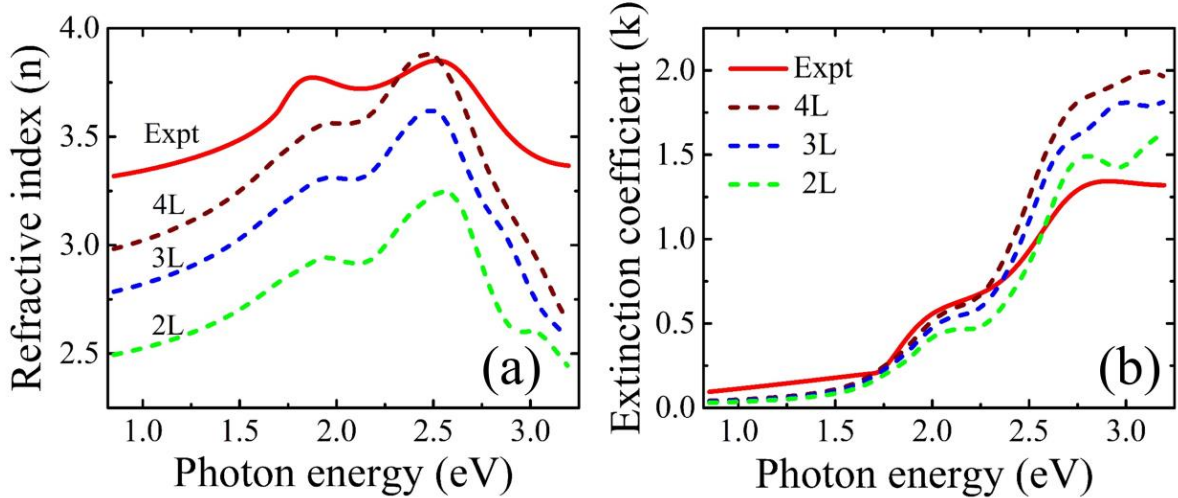

*Figure 6. Real (a) and imaginary (b) part of complex refractive index. Red solid curve: experimental spectrum of the FL-MoS$_2$ sample; dashed curve: theoretical a-b plane data for 2,3, and 4 layer MoS$_2$*

Kronig consistent Tauc-lorentz (T-L) oscillators [32]. For the fit shown in Fig. 5, a thickness of 2.6 nm was obtained (see Table S4 for more details on fit). The data for all angles could be fit using the same model and thickness values.

In Figure 6 we plot the real and imaginary part of the complex refractive index (*n* and *k*) for the FL-MoS$_2$ sample. The calculated *ab*-plane *n* and *k* values for MoS$_2$ layers of different thicknesses is shown as a guide and to help identify prominent optical transitions. As is clear from a casual inspection, the best agreement with our experimental refractive index value is with the calculated four-layer data (Fig. 6a). The strong features centered at 1.87 and 2.52 eV in the experimental refractive index is very close to the theoretical maxima at 1.94 and 2.56 eV respectively, as highlighted. Numerous works have identified these transitions as A/B peak and C peak [33, 34]; A and B peaks are related to the spin-orbit split valance electrons directly transit to the VB minimum at K (K') points. C peak is identified as the electron transition from valance band to conduction band at the wave vectors lie in between Gamma and K points. The latter is identified as a result from the band nesting effect [35, 36]. Results from our ab-initio DFT calculations without considering the excitonic effects correctly identifies the major peaks in the experimental optical spectra [20, 33, 34]. Likewise, the features in the extinction coefficient match between theory and experiment very well. The variation of the experimental optical constants with thickness and annealing are shown and discussed in Fig S5-S7.

We have further studied the electrical properties of the $MoS_2$/BN heterostructures. Current-Voltage (I-V) measurements with two probe and four probe configurations were measured on the deposited films. For the two-probe measurements Chromium (Cr) and Gold (Au) contact pads were deposited using thermal evaporation through a metal shadow mask, and standard lithography process was followed in order to deposit four contacts using Titanium (Ti) and Gold (Au). For the two-probe measurements samples were mounted on the chip holder inside the closed cycle Helium cryostat (Janis Model # SHI- 4-1) and were pumped down to high vacuum level (~$10^{-6}$ Torr) before performing the electrical measurements. The four- probe measurements were performed under ambient conditions. Room temperature current voltage (I-V) measurements were performed using Keithley source meters (2400 series). For the I-V measurements, samples were forward and reverse biased by applying ±1V between the contacting electrodes. It is well known that any metal semiconductor junction give rise to Schottky barrier and it is very difficult to circumvent the Schottky barrier between the semiconducting channel and the metallic contacts. However, the linearity of IV curves (which is a lot more prominent at low applied voltages) signifies that within the low applied voltage regime, the contacts are able to replenish the charge carriers efficiently when they are drawn out from the material under an applied electric field. Under such conditions we can assume that the contacts are "behaving" as "ohmic" rather than blocking or injecting.

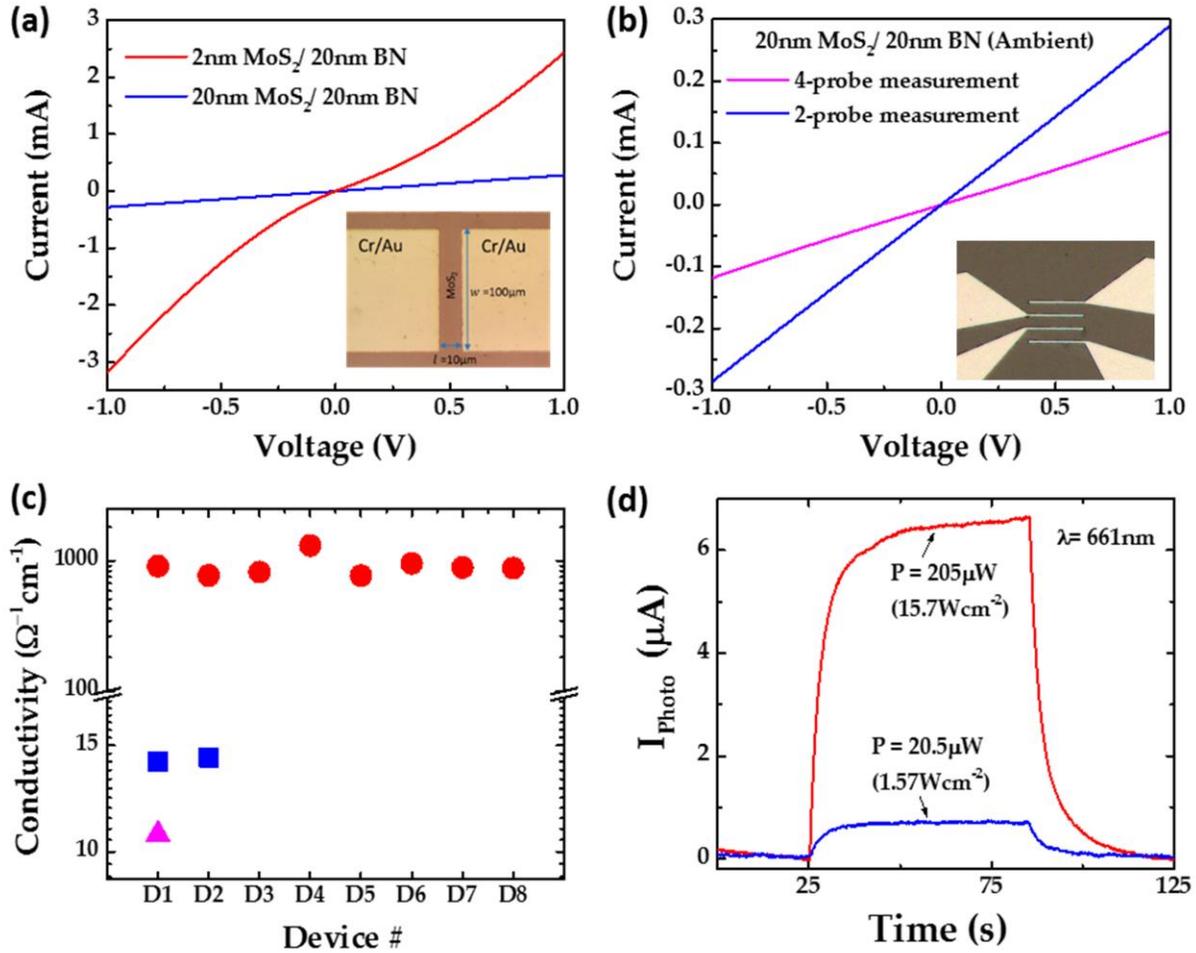

*Figure 7 (a) Current-voltage (I-V) response obtained from 2 nm (red) & 20 nm MoS2 (blue) devices under high vacuum condition in 2-probe geometry (inset) Optical image of a device with Cr/Au contact pads with 2-probe device geometry (l = 10μm, w = 100μm). (b) Current-voltage (I-V) response obtained for 20nm MoS2/BN devices with 2-probe (blue)) & 4-probe (pink) geometry under ambient conditions (inset) Optical image of a device with Au contact pads with 4-probe device geometry (l = 15μm, w = 87μm). (c) Conductivity measurements of several 2 nm MoS2 (red circles), 20nm MoS2 (blue squares) in 2-probe geometry & 20nm MoS2 (pink triangle) in 4-probe geometry. (d) Photo current at different laser illumination power ($V_{SD}$=2V, λ=661nm) for the 20 nm MoS$_2$ sample.*

Figure 7 (a) shows the I-V curves of both the few-layer MoS$_2$ (2nm)/BN and many-layer MoS$_2$ (20nm)/BN heterostructures. Strikingly, 2 nm MoS$_2$/BN devices shows lower resistance (~500Ω) than the 20nm MoS$_2$/BN devices (~3.5kΩ). In figure 7 (b) a comparison of the IV measurement on 20nm MoS$_2$ measured with two point contact and four point contact is shown.

Further we found the conductivity of several devices for these heterostructures. Conductivity ($\sigma$) is defined as $\sigma = l/R.w.t$, where $l$ is the length between two electrode pads, $w$ is the width of an electrode, $t$ is the height of the MoS$_2$ layers and $R$ is the measured resistance of the heterostructure devices. We found that conductivity of 2 nm MoS$_2$/BN heterostructure is ~2 orders higher than the 20nm MoS$_2$/BN heterostructure (Figure 7 (c)). Also the conductivity of both of these heterostructures were several orders higher than the bulk like MoS$_2$ (~$10^{-8}$ $\Omega^{-1}$cm$^{-1}$) [37]. Here we would like to note that the measured conductivity with two point contact and four point contact on 20 nm MoS$_2$/BN device are of similar order. This is indicative of the fact that the two point measurements which were measure using larger contact pad, perhaps provides less contact resistance and hence is similar to the four point measurement where the four point geometry minimizes the contact resistance.

In the past several other groups have investigated the electrical properties of MoS$_2$ grown or obtained using several other techniques. For example Linear conductivity ($\sigma'$) of mechanically exfoliated MoS$_2$ devices [38] and CVD grown MoS$_2$ [39] devices are indicated in Table 1. Where $\sigma'$ is defined as $\sigma' = l/R.w$. To compare with the literature we also calculated the linear conductivity of sputter deposited MoS$_2$ devices. As indicated in the Table 1, $\sigma'$ of sputter deposited MoS$_2$ devices are several orders higher than the mechanically exfoliated MoS$_2$ devices as well as CVD grown MoS$_2$ devices.

Here we note that the increased conductivity of few-layer films (figure 7c and Table 1) seems contrary to the scaling behavior observed in conventional (3D) systems. First of all, in order to confirm that the conductivity in our samples is indeed from the MoS$_2$ layer, and not leaking from the conducting p-doped Si substrate beneath the 20 nm a-BN buffer layer we have conducted additional photo-current measurements on a 20nm BN layer deposited on Silicon substrate (see supplementary figure S8) . As the figure indicates, the response of bare BN devices on Si is insensitive to incident light. We therefore found no evidence of substrate-assisted photo-conduction. This also supports the conclusion that the substrate cannot explain increase of conductivity in few-layer samples.

Table 1. Linear Conductivity values of $MoS_2$ devices developed by different techniques

| Device | Thickness | Linear Conductivity ($\Omega^{-1}$) | Reference |
|---|---|---|---|
| **Mechanically exfoliated $MoS_2$** | 1 Layer | ~1 x $10^{-8}$ | [38] |
| | 3 Layers | ~2.5 x $10^{-7}$ | [38] |
| **CVD grown $MoS_2$** | 1-4 Layers | ~3.52 x $10^{-5}$ - 6.85 x $10^{-5}$ | [39] |
| **Sputter deposited $MoS_2$** | 2 nm (2-probe Vac.) | ~1.8 x $10^{-4}$ | This work |
| | 20nm (2-probe Vac.) | ~2.82 x $10^{-5}$ | This work |
| | 20nm (2-probe Amb.) | ~2.86 x $10^{-5}$ | This work |
| | 20nm (4-probe Amb.) | ~2.00 x $10^{-5}$ | This work |

We can also put forward a few explanations as to why the enhanced conductivity of few-layer $MoS_2$ is reasonable. Firstly, Raman and XRD characterization (Fig 3 and 4) clearly indicates that sputter-deposited many-layer films are more disordered compared to few-layer films that can explain increased conductivity. Secondly, there is also literature evidence supporting inverse-thickness scaling behavior in TMD thin films [41] which is also consistent with our results. Additionally, it is also well-documented that structural modification from 2H to 1T polytype within the $MoS_2$ layer can lead to metallic behavior [42-43]. Therefore, all such evidence taken together clearly dispels the notion that conductivity is not necessarily more in case of thicker films.

In order to gauge the potential application of these films we have performed preliminary photoconductivity measurement of the 20 $MoS_2$/BN device with four-point contact (figure 7 (d)). Room temperature photo switching behavior of the 20nm $MoS_2$/BN devices were investigated by illuminating a λ=661nm laser light source with two different illumination intensities. 2V bias voltage was applied in-between source and drain terminals. In order to calculate the photo current, current passing through the device without light illumination (Dark Current; $I_d$) and with light illumination (Illuminated current; $I_{ill}$) was recorded. The laser was switched on and off for about 60s at a constant intensity. This sequence was followed for 1.57 Wcm$^{-2}$ and 15.7 Wcm$^{-2}$ laser intensities. The photo current ($I_{ph}$) at a particular intensity was calculated by using the equation $I_{ph}$ = $I_{ill}$ - $I_d$. The photo switching behavior at two different laser intensities are shown in Figure 7(d). Despite the fact that photocurrent of higher laser intensity was an order higher than the photocurrent of lower laser intensity, Responsivity (R) values were remained to be around ~31 mAW$^{-1}$ for both the laser intensities. Responsivity is one of the important performance parameter of photoactive materials, is defined as the ratio of the $I_{ph}$ and the laser illumination intensity ($P_{light}$). The responsivity data obtained is compared with photo response of other $MoS_2$ based materials (See Table S6 in supplementary information). We also note that the R values of the MoS2/BN in some cases are higher than the R values of single/few layer $MoS_2$ devices at similar growth/ measurement condition. These preliminary measurements indicate the possible use of these films in various large area opto-electronic applications.

## CONCLUSION

In conclusion, we have successfully grown and investigated the structural, optical, and transport properties of few-layer $MoS_2$ (down to 4 monolayers) film grown on amorphous Boron Nitride buffered silicon substrates. A room or low temperature growth followed by post deposition annealing at a higher temperature provides the best quality $MoS_2$ layer with sharp interfaces. Growth of ordered few-layer crystalline $MoS_2$ is confirmed by X-ray, Raman spectroscopy, and ellipsometry measurements. Growth of ordered few-layer crystalline $MoS_2$ is confirmed by X-ray, Raman spectroscopy, and ellipsometry measurements. Furthermore, the few-layers films show lower disorder compared to many-layer films. Optical properties of the $MoS_2$ few-layer is shown to be in very good agreement with band structure calculations. : Transport measurements point to the conclusion that few-layer $MoS_2$ films have better conductivity than many-layer sputtered films, and compare favorably with other large-area growth methods such as CVD. Further, we have also recorded significant photo responsive behavior of these films. This has interesting and positive implications for device applications for large-area TMD films grown using sputtering.

## ACKNOWLEDGEMENTS

DM would like to thank startup funds at Southern Illinois University and ST would like to acknowledge the support provided by the U.S. Army Research Office through a MURI grant # W911NF-11-1-0362. The spectroscopic ellipsometry data shown here was carried out in the Frederick Seitz Materials Research Laboratory Central Research Facilities, University of Illinois. D.M. would like to acknowledge Dr. Julio Soares for his help and guidance with Ellipsometry and Raman measurements.